\documentclass[aoas,nameyear,dvips]{arximspdf}
\usepackage{algorithm,algorithmic}
\usepackage{dcolumn}
\usepackage{graphicx}

\doi{10.1214/09-AOAS299}
\volume{4}
\issue{2}
\pubyear{2010}
\firstpage{830}
\lastpage{848}

\makeatletter

\newcolumntype{d}[1]{D{.}{.}{#1}}
\makeatother

\begin{document}
\begin{frontmatter}

\title{Likelihood inference for particle location in fluorescence microscopy\protect\thanksref{T1}}
\runtitle{Particle location in fluorescence microscopy}
\thankstext{T1}{Supported by the NSF/NIH joint initiative in
mathematical biology DMS-07-14939.}

\begin{aug}
\author[a]{\fnms{John} \snm{Hughes}\ead[label=e1]{jph264@psu.edu}},
\author[a]{\fnms{John} \snm{Fricks}\ead[label=e2]{fricks@stat.psu.edu}\corref{}}
\and
\author[b]{\fnms{William} \snm{Hancock}\ead[label=e3]{wohbio@engr.psu.edu}}
\runauthor{J. Hughes, J. Fricks and W. Hancock}
\affiliation{Pennsylvania State University}
\address[a]{J.Hughes\\ J. Fricks\\
Department of Statistics\\
Pennsylvania State University\\
University Park, Pennsylvania 16802\\USA\\
\printead{e1}\\
\phantom{E-mail: }\printead*{e2}
}
\address[b]{W. Hancock\\ Department of Bioengineering\\
Pennsylvania State University\\
University Park, Pennsylvania 16802\\USA\\
\printead{e3}
}
\end{aug}

% HISTORY:
\received{\smonth{12} \syear{2008}}
\revised{\smonth{10} \syear{2009}}

% ABSTRACT
%
\begin{abstract}
We introduce a procedure to automatically count and locate the
fluorescent particles in a microscopy image.
Our procedure employs an approximate likelihood estimator derived from
a Poisson random field model for
photon emission. Estimates of standard errors are generated for each
image along with the parameter
estimates, and the number of particles in the image is determined using
an information criterion and
likelihood ratio tests. Realistic simulations show that our procedure
is robust and that it leads to accurate
estimates, both of parameters and of standard errors. This approach
improves on previous {\it ad hoc} least
squares procedures by giving a more explicit stochastic model
for certain fluorescence images and by employing a consistent framework
for analysis.
\end{abstract}

% KEYWORDS
%
\begin{keyword}
\kwd{Maximum likelihood methods}
\kwd{Poisson random field}
\kwd{fluorescence microscopy}
\kwd{particle tracking}
\kwd{organelle}
\kwd{molecular motor}
\kwd{nanotechnology}.
\end{keyword}

\end{frontmatter}

%s1 ###
\section{Introduction}
\label{sec:intro}

The accurate and precise tracking of microscopic fluorescent particles
attached to biological specimens (e.g., organelles, membrane proteins,
molecular motors) can give insights into the nanoscale function and
dynamics of those specimens. This tracking is accomplished by analyzing
digital images produced by a CCD (charge-coupled device) camera
attached to a microscope used to observe the specimens repeatedly. In
this paper we introduce an improved technique for analyzing such images
over time. Our method, which applies maximum likelihood principles,
improves the fit to the data, derives accurate standard errors from the
data with minimal computation, and uses model-selection criteria to
``count'' the fluorophores in an image. The ability to automate the
process and quickly derive standard errors should allow for the
analysis of thousands of images obtained from a typical experiment and
aid in methods to track individual fluorophores across sequential images.

In fluorescence microscopy, a specimen of interest is tagged with a
fluorescent molecule or particle. The fluorescence microscope then
irradiates the specimen with light at the excitation wavelength of the
fluorophore, and when the excited electrons revert to the ground state
they emit photons at the emission wavelength. A filter separates the
emitted light from the excitation light so that only the light from the
fluorescent material can pass to the microscope's eyepiece and camera
system [\citet{Rost1992Fluorescence-Mi}].

In general, the Rayleigh criterion implies that the maximum resolution
for a light microscope should be roughly 250 nm (half of the wavelength
of visible light); however, Selvin and his collaborators found that by
fitting the center point of the point spread function one can locate a
particle of interest. This technique is known as FIONA (Fluorescence
Imaging with One-Nanometer Accuracy) and was introduced in \citet
{Yildiz2003Myosin-V-Walks-}. The key element of FIONA is to focus
attention on single fluorophores used as markers in biological
specimens [\citet{0953-8984-17-47-023}]. By analyzing sequences of
images, molecular motors (e.g., myosin VI and kinesin) and other
specimens can be tracked through time, giving researchers insight into
their dynamics and biological function. For instance, Yildiz et al.
used FIONA to find compelling support for the hypothesis that myosin V
walks hand over hand and evidence to eliminate other hypotheses [\citet
{Yildiz2003Myosin-V-Walks-}; \citet{0953-8984-17-47-023}].

A number of analysis techniques have been proposed for FIONA\break images. In
2001, \citeauthor{Cheezum2001Quantitative-Co} compared four\break
methods---cross-correlation, sum-absolute difference, centroid, and
Gaussian fit---and ultimately recommended the Gaussian-fit method for
single-fluorophore tracking. In the Gaussian-fit approach, the method
of ordinary least squares (OLS) is used to fit a sum of symmetric
bivariate Gaussian functions to the image. Least squares fitting is
relatively efficient, and software to do it is widely available.
Thompson, Larson and Webb subsequently proposed a ``Gaussian mask''
algorithm that is easier to implement than the Gaussian-fit method, is
computationally less intensive, and performs nearly as well in
simulations (\citeyear{Thompson2002Precise-Nanomet}). The
Gaussian-mask algorithm is essentially a centroid calculation that
weights each pixel with the number of photons in the pixel and with a
bivariate Gaussian function integrated over the pixel. In both cases,
simulation studies using typical experimental values showed that
sub-pixel or even nanometer resolution was possible.

The above mentioned Gaussian-fit and Gaussian-mask methods, while
appealing, share two shortcomings. Since one or more beads may move out
of frame for a particular image, the number of beads from one image to
the next is not known {a priori} and must be determined for each
image. Previous authors have attempted to solve this problem by means
of a grid search, the first step of which is to scan the image for all
pixels greater than some arbitrary threshold value. Each of these
extreme pixels is taken to be a bead location, and some region
surrounding each extreme pixel is extracted from the image and
processed by OLS or the Gaussian mask. Thompson et al. suggest a
threshold that is eight standard deviations above the mean pixel value,
but no explicit evidence is given in support of this choice [\citet
{Thompson2002Precise-Nanomet}]. The correct threshold level for a set
of images could possibly be approximated using simulations, but this
would be a complex and computationally intensive task that would be
necessary for each set of images, since the level of background noise
may vary significantly from one experiment to the next.

The second drawback is that estimates of precision are derived from
simulation studies alone. If the probability model for the problem is
misspecified, then error estimates based on simulations from that model
will be inaccurate even if reasonable parameter values were used in the
studies. And these values will vary from image to image due to changing
experimental conditions, for example, elevated background noise or
slight changes in focus. A possible solution is to perform a Monte
Carlo simulation study using parameter values derived from the current
experiment. But, given that the fitting procedures themselves are time
consuming, these approaches to standard error calculation may prove
infeasible. It takes our algorithm several minutes---on a dual 2.8GHz
Quad-Core Intel Xeon Mac Pro---to process an image with fifteen
particles, which implies that bootstrapping standard errors for such an
image would require hours of computation. Moreover, a full analysis of
an experiment requires processing many hundreds of images.

In what follows we present a new approach to counting and locating
fluorophores. Our approach eliminates the need for a grid search and
estimates standard errors from the data, without additional simulation,
via standard likelihood tools. In Section~\ref{sec:mod} we present an
explicit probability model for a FIONA image along with a maximum
likelihood estimation procedure suitable for this model. In Section~\ref
{sec:se} we discuss the properties of the approximate likelihood
estimator presented in Section~\ref{sec:mod}. In Section~\ref{sec:sel}
we discuss stepwise model selection, which allows our procedure to
automatically determine the number of beads in an image in a consistent
manner. In Section~\ref{sec:sim} we describe the results of realistic
simulation studies that support the approximations presented in
Section~\ref{sec:mod} and demonstrate the robustness of our procedure.
Finally, in Section~\ref{sec:ex} we carry out a complete analysis of an
experimentally collected FIONA image, introducing relevant diagnostic
criteria for our fit.

%s2 ###
\section{Model for data from a single FIONA image}
\label{sec:mod}

We first develop a model for the photon emission from particles
distributed over a microscope slide. Photon emission from a constant
source generally follows a Poisson distribution; this fact naturally
leads to a Poisson random field model for the emission from a slide. We
then express the effect of pixelation on the field, representing the
emission as viewed from the digital camera, and employ a normal
approximation to the Poisson distribution. We arrive at our final
approximate model by accounting for additional error introduced by the
camera and its associated equipment.

We begin with the standard model for the photon-emission pattern (as
distorted by the point-spread function of a microscope objective) of a
collection of fluorophores distributed at random over some region of
$\mathbb{R}^2$ [\citet{Cheezum2001Quantitative-Co}; \citet{Thompson2002Precise-Nanomet}].

Let $N$, a Poisson random field on a rectangular subset $T$ of $\mathbb
{R}^2$, represent the emission pattern of the sample. The intensity
function can be defined for any Borel set $R\subset T$ as
%
%e1 ###
\begin{equation}
\ \,E\{N(R)\}=\int\hspace*{-3pt}\int_R\Biggl\{B+\sum\limits_{j=0}^{J-1}A_j\cdot\exp\biggl(-\frac
{(x-x_j)^2+(y-y_j)^2}{S^2}\biggr)\Biggr\}\,dx\, dy.
\end{equation}
Thus, $N(R)$ is a Poisson random variable with the mean equal to a sum
of $J$ Gaussian functions, one for each bead, with Gaussian function
$j$ symmetric about $(x_j,y_j)$ (which is contained in $T$). In
addition, there is a constant background intensity of magnitude $B$
representing background fluorescence. Although the intensity function
for a bead is more often modeled in the physics literature by an Airy
function, a Gaussian function approximates the Airy function quite
well, and so we take the Gaussian centered at $(x_j,y_j)$ to represent
the (distorted) emission of bead $j$ [\citet{Saxton1997Single-particle};
\citet{Thompson2002Precise-Nanomet}].

The photons emitted by the sample are collected by a camera, the pixels
of which can be represented by partitioning $T$ into a uniform grid,
where each pixel in the grid is square with side length $(a)$ nm. Then,
for a given pixel $Z_i$ with center $(x_i,y_i)$,
\begin{eqnarray}
\quad E(Z_i)&=&\int_{y_i-a/2}^{y_i+a/2}\int_{x_i-a/2}^{x_i+a/2}
\Biggl\{B+\sum_{j=0}^{J-1}A_j\nonumber
\\[-8pt]\\[-8pt]
&&{}\hspace*{120pt}\cdot\exp\biggl(-\frac{(x-x_j)^2+(y-y_j)^2}{S^2}\biggr)\Biggr\}\,dx \,dy,\nonumber
\end{eqnarray}
which is approximately equal to
%
%e2 ###
\begin{equation}
\biggl\{B+\sum\limits_jA_j\cdot\exp\biggl(-\frac{(x_i-x_j)^2+(y_i-y_j)^2}{S^2}\biggr)\biggr\}a^2.
\end{equation}
Since $a^2$ is a constant, we allow the $A_j$ and $B$ to absorb it,
arriving at
%
%e3 ###
\begin{equation}
E(Z_i)\approx B+\sum_jA_j\cdot\exp\biggl(-\frac
{(x_i-x_j)^2+(y_i-y_j)^2}{S^2}\biggr)=f_i.
\end{equation}

Moreover, $B$ is generally large enough to justify using a normal
approximation to the Poisson distribution. More precisely,
%
%e4 ###
\begin{equation}
Z_i\stackrel{\cdot}{\sim}N(f_i,f_i),
\end{equation}
where we note that $f_i$ depends on the parameters of interest. In this
model it is obviously important that $B$ and the $A_j$'s be constrained
so that $f_i$ is nonnegative.

The discretized Poisson random field described above is taken as the
underlying model for the photon emission; however, additional error,
which we will call instrumentation error, arises from various sources
such as signal quantization and dark current, an electric current that
flows through a CCD even when no light is entering the device [\citet
{bobroff1152}; \citet{Thompson2002Precise-Nanomet}]. If we model the
instrumentation error as a $N(0,\theta)$ random variable independent of
the intensity, then we have as a final approximate model for the data
%
%e5 ###
\begin{equation}
Z_i\stackrel{\cdot}{\sim}N(f_i,f_i+\theta).
\end{equation}

Consequently, the approximate likelihood of a given image with $n$
pixels is
%
%e6 ###
\begin{equation}
L_n(\bolds \beta)=\prod_{i=1}^n\frac{1}{\sqrt{2\pi(f_i+\theta
)}}\exp\biggl(-\frac{(z_i-f_i)^2}{2(f_i+\theta)}\biggr),
\end{equation}
where $\bolds \beta=(x_0,y_0,A_0,\ldots
,x_{J-1},y_{J-1},A_{J-1},S,B,\theta)^T$, the parameters of interest.
This implies that the log-likelihood, without unnecessary constants, is
%
%e7 ###
\begin{equation}
\ell_n(\bolds \beta)=-\sum_i\ln(f_i+\theta)-\sum
_i\frac{(z_i-f_i)^2}{f_i+\theta},
\end{equation}
which we maximize with respect to $\bolds \beta$ to obtain $\hat
{\bolds \beta}_{\mathrm{MLE}}.$

%s3 ###
\section{Estimating standard errors}
\label{sec:se}

From the theory of maximum likelihood estimators we know that, provided
certain regularity conditions are met, a properly scaled MLE converges
asymptotically to a normally distributed random variable. In our case,
%
%e8 ###
\begin{equation}
[\bolds{\mathcal{I}}_n(\bolds \beta)]^{1/2}(\hat{\bolds
\beta}_n-\bolds \beta)\stackrel{D}{\longrightarrow}N_p(\bolds
0,\mathbf{I}_p),
\end{equation}
where $\bolds{\mathcal{I}}_n(\bolds \beta)$ is the Fisher
information about $\bolds \beta$ contained in a sample of size $n$,
$p$ is the dimension of $\bolds \beta$, and $\mathbf{I}$ is the $p
\times p$ identity matrix. This implies that
%
%e9 ###
\begin{equation}
\hat{\bolds \beta}_n\stackrel{\cdot}{\sim}N_p(\bolds \beta
,[\bolds{\mathcal{I}}_n(\bolds \beta)]^{-1})
\end{equation}
for large $n$, and so the diagonal elements of $[\bolds{\mathcal
{I}}_n(\bolds \beta)]^{-1}$ are approximate sampling variances for
the estimators $\hat{\bolds \beta}_n$. However, we do not know
$[\bolds{\mathcal{I}}_n(\bolds \beta)]^{-1}$ because the true
$\bolds \beta$ is unknown and analytical calculation of the
information is prohibitively complicated.

Consequently, we use the standard substitution $[\bolds{\mathcal
{I}}_n(\bolds \beta)]^{-1}$ using the observed information,
$\bolds{\mathcal{J}}_n(\hat{\bolds \beta}_n)=[-\frac{\partial
^2}{\partial\bolds{\beta}_i\,\partial\bolds{\beta}_j}\ell_n(\hat
{\bolds\beta}_n)]$, that is, $
[\bolds{\mathcal{I}}_n(\bolds \beta)]^{-1}\approx\break[\bolds
{\mathcal{J}}_n(\hat{\bolds \beta}_n)]^{-1}
$. Estimating $[\bolds{\mathcal{I}}_n(\bolds \beta)]^{-1}$ by
inverting $\bolds{\mathcal{J}}_n(\hat{\bolds \beta}_n)$ has
the advantage of avoiding closed form derivatives which are unwieldy in
this case.

% $\hat{x}_j\stackrel{\cdot}{\sim}N[x_j,\widehat{Var}(\hat{x}_j)]$ and
%$\hat{y}_j\stackrel{\cdot}{\sim}N[y_j,\widehat{Var}(\hat{y}_j)]$,
%where $\widehat{Var}(\hat{x}_j)$ and $\widehat{Var}(\hat{y}_j)$
%represent the diagonal elements of $$ corresponding to $\hat{x}_j$ and
%$\hat{y}_j$, respectively. Moreover,
The preceding implies that the joint distribution of $\hat{x}_j$ and
$\hat{y}_j$ is approximately bivariate normal. More precisely,
%
%e10 ###
\begin{equation}
\hat{\bolds \mu}_j=(\hat{x}_j,\hat{y}_j)^T\stackrel{\cdot}{\sim
}N_2[\bolds \mu_j=(x_j,y_j)^T,\widehat{\bolds \Sigma}_j],
\end{equation}
where
%
%e11 ###
\begin{equation}
\widehat{\bolds \Sigma}_j=[\bolds{\mathcal{J}}_n(\hat
{\bolds \beta}_n)]^{-1}=
\left[
\matrix{\widehat{\operatorname{Var}}(\hat{x}_j)&\widehat{\operatorname{Cov}}(\hat{x}_j,\hat
{y}_j)
\cr
\widehat{\operatorname{Cov}}(\hat{x}_j,\hat{y}_j)&\widehat{\operatorname{Var}}(\hat{y}_j)
}
\right]
.
\end{equation}

Since the contours, that is, the equidensity curves, of the bivariate
normal distribution are ellipses, an approximate 95\% confidence region
for the location of bead $j$ also takes the form of an ellipse:
%
%e12 ###
\begin{equation}
\{ \bolds \mu_j\dvtx (\hat{\bolds \mu}_j-\bolds \mu_j)^T\widehat
{\bolds \Sigma}^{-1}_j(\hat{\bolds \mu}_j-\bolds \mu_j)=\chi
^2_{0.95,2} \}
,
\end{equation}
where $\chi^2_{0.95,2}$ denotes the 95th percentile of the $\chi^2$
distribution with 2 degrees of freedom [\citet
{Ravishanker2002A-First-Course-}]. For an image with multiple beads,
the typical case, we may want a collection of random ellipses, the
union of which will enclose all $J$ beads with probability 0.95. We can
accomplish this by using the Bonferroni correction, which assigns to
each bead an error rate of $\frac{0.05}{J}$, thereby making the image
wide error rate 0.05. The resulting collection of simultaneous
confidence ellipses is given by
%
%e13 ###
\begin{equation}
\{ \bolds \mu_j\dvtx (\hat{\bolds \mu}_j-\bolds \mu_j)^T\widehat
{\bolds \Sigma}^{-1}_j(\hat{\bolds \mu}_j-\bolds \mu_j)=\chi
^2_{1-0.05/J,2} \}.
\end{equation}

We evaluated our standard-error estimation and the convergence of our
estimator by way of a simulation study. Ten thousand $100\times100$
single-bead images were simulated, each image having its lone bead
located at $(7823,3353)$, where the coordinates are given in nanometers
from the lower left corner. (For an idea about the nature of the data,
see Figure \ref{fig:dim}, which has four beads.) Table~\ref{table:se}
shows $\hat{\bolds \beta}_{\mathrm{MLE}}$ for a single image along with
standard-error estimates for that image and the true standard errors
gleaned from all 10,000 images. Our estimated standard errors are in
close agreement with the true standard errors.

%t1 ###
\begin{table}[b]
\tablewidth=220pt
\caption{Estimation of standard errors. Estimation error is the true
value of the parameter minus the estimated value}\label{table:se}
\begin{tabular*}{220pt}{@{\extracolsep{4in minus 4in}}ld{6.0}d{2.1}d{2.3}d{2.3}@{}}
\hline
\textbf{Parameter}&\multicolumn{1}{c}{\textbf{Truth}}&\multicolumn{1}{c}{\textbf{Est error}}&
\multicolumn{1}{c}{\textbf{Sim SE}}&\multicolumn{1}{c@{}}{$\widehat{\mathbf{SE}}$}\\
\hline
$x_0$ & 7823 & -0.4 & 0.420 & 0.422 \\
$y_0$ & 3353 & 0.1 & 0.424 & 0.421\\
$A_0$ & 15{,}000 & 99.8 & 77.4& 61.2\\
$S$ & 200 & -0.2 & 0.393& 0.341\\
$B$ & 200 & 0.1 & 0.173 & 0.171 \\
$\theta$ & 100 & 6.7 & 4.26 & 4.16\\
\hline
\end{tabular*}
\end{table}

\begin{figure}

\includegraphics{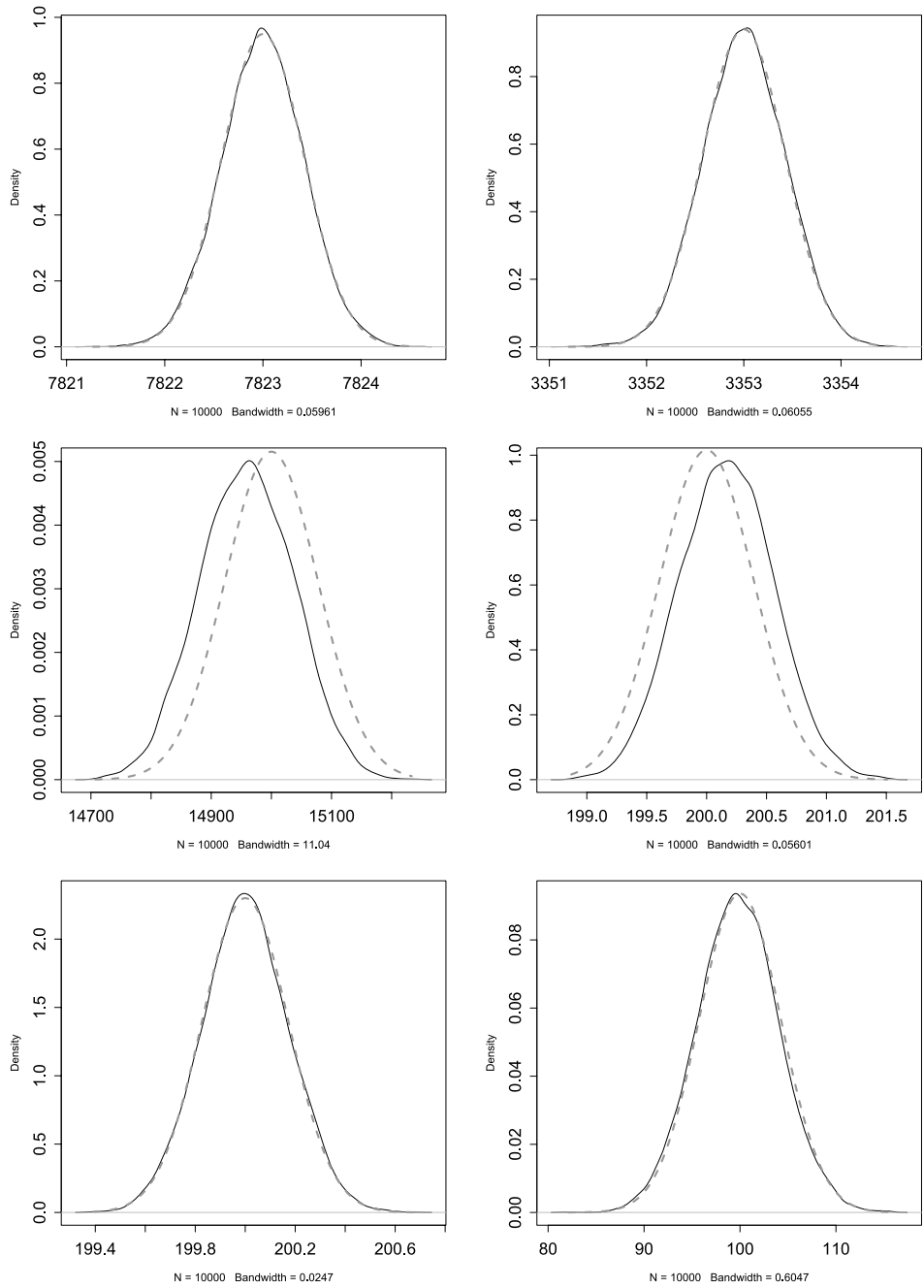}

\caption{These plots show density estimates for the sampling
distributions of $\hat{x}_0$, $\hat{y}_0$, $\hat{A}_0$, $\hat{S}$, $\hat
{B}$, and $\hat{\theta}$, respectively, with normal densities (dashed)
superimposed.}
\label{fig:sampdist}
\end{figure}

Figure~\ref{fig:sampdist} shows estimated densities for the sampling
distributions of $\hat{x}_0$, $\hat{y}_0$, $\hat{A}_0$, $\hat{S}$, $\hat
{B}$, and $\hat{\theta}$, respectively. Appropriate normal densities
(dashed) are shown superimposed. Each normal density is centered at the
true value for its parameter. It is clear that the sampling
distributions converge to normality, but the estimators of $S$ and
$A_0$ are slightly biased in opposing directions; intuitively,
estimation of $S$ works in opposition to that of $A_0$. This is because
the fitting procedure is attempting to simultaneously conform $\hat{S}$
to the base of the Gaussian peak and $\hat{A}_0$ to the peak's height,
and an adjustment of either estimate nudges the other in the opposite direction.

%f1 ###

%s4 ###
\section{Model selection: how many beads are in the image}
\label{sec:sel}

Previous authors have suggested that the number of beads in an image be
determined by applying a grid search algorithm prior to fitting [\citet
{Cheezum2001Quantitative-Co}; \citet{Thompson2002Precise-Nanomet}]. As we
mentioned in Section \ref{sec:intro}, any pixel with an intensity above
some threshold is identified with a bead, and then some region in the
vicinity of the pixel is extracted from the image and fitted. This
thresholding approach may be adequate for producing initial estimates
of bead locations, but thresholding prevents full automation because
the threshold must be chosen by the investigator. And even a seemingly
well-chosen threshold may be too large to distinguish dim beads from
background noise.

Our procedure eliminates these problems by fitting first and selecting
the number of beads based on those fits. In our scheme, an approximate
information criterion derived from an OLS fit is used to approximate
the size of the model, and then, because the candidate models are
nested, likelihood ratio tests are used to select the final model. As
we will show in Section \ref{sec:sim}, this approach is able to
identify all of the beads, even very dim ones, automatically.

Our algorithm has a preliminary stage for estimating the number of
beads and producing initial estimates of all parameters except $\theta
$, and a final stage for estimating $\theta$, giving maximum likelihood
estimates of the other parameters, and accurately determining the
number of beads. The preliminary stage assumes zero beads initially and
fits $f(x,y)=B$ to the image using OLS. Using the least squares fit at
each stage, the information criterion
%
%e14 ###
\begin{equation}
\mathit{IC}^{(k)}=n\ln\biggl(\frac{\mathit{RSS}}{n}\biggr)+p \sqrt{n}
\end{equation}
is computed, where $k$ is the (assumed) number of beads, $n$ is the
sample size, $\mathit{RSS}$ is the residual sum of squares, and $p$ is the
number of free parameters. Note that $\mathit{IC}$ is an increasing function of
$\mathit{RSS}$ and $p$, which implies that $\mathit{IC}$ rewards a better fit (smaller
$\mathit{RSS}$) and penalizes more free parameters. On the next iteration, one
bead is assumed, and so
%
%e15 ###
\begin{equation}
f(x,y)=B+A_0\cdot\exp\biggl(-\frac{(x-x_0)^2+(y-y_0)^2}{S^2}\biggr)
\end{equation}
is fit to the image, producing $\mathit{IC}^{(1)}$. Iteration continues until
$\mathit{IC}^{(k)}>\mathit{IC}^{(k-1)}$, which indicates that the image contains $k-1$
beads.

Note that $\mathit{IC}$ is a nonstandard information criterion. We found that
even the Bayesian Information Criteria (BIC) does not penalize
additional parameters sufficiently. This can allow the initial stage of
the algorithm to significantly overestimate the correct number of
beads, causing much unnecessary computation during the final stage of
the algorithm. Our simulations showed that replacing BIC's $\ln(n)$
with $\sqrt{n}$ minimizes overfitting.

As the algorithm makes an initial forward sweep over possible models,
the OLS parameter estimates are saved. This initial sweep stops based
on the information criteria, $\mathit{IC}$. Those estimates are used to
initialize the maximum likelihood estimation carried out in the final
backward sweep which terminates using likelihood ratio criteria.
Providing the MLE routine with good initial estimates of all parameters
except $\theta$ allows the MLE to converge faster than it otherwise would.

The parameter fits and the selection criteria initially computed are
then used to find the final parameter estimates and make the final
model selection. The key differences are that OLS is replaced by MLE
and $\mathit{IC}$ is replaced by the likelihood ratio statistic
%
%e16 ###
\begin{equation}
G^2_{(\mathit{beads})}=-2\bigl\{\ell_n\bigl(\hat{\bolds \beta}_{\mathrm{MLE}}^{(\mathit{beads})}\bigr)-\ell
_n\bigl(\hat{\bolds \beta}_{\mathrm{MLE}}^{(\mathit{beads}+1)}\bigr)\bigr\},
\end{equation}
which should be approximately $\chi^2$ distributed with three degrees
of freedom (because each bead is associated with three parameters:
$A_j$, $x_j$, and $y_j$). The full algorithm is given in pseudocode below.

%$\mathit{beads} \leftarrow 0$
%}
%}_{\mathrm{MLE}}^{(\mathit{beads})})\} \leftarrow\operatorname{MLE}{\mathit{pixels},\mathit{beads}, \hat{\bolds
%}_{\mathrm{MLE}}^{(\mathit{beads})})\} \leftarrow\operatorname{MLE}{\mathit{pixels},\mathit{beads}, \hat{\bolds
%G^2_{(\mathit{beads})} \leftarrow-2\{\ell_n(\hat{\bolds \beta
%}_{\mathrm{MLE}}^{(\mathit{beads})})-\ell_n(\hat{\bolds \beta}_{\mathrm{MLE}}^{(\mathit{beads}+1)})\}
%}

\renewcommand{\thealgorithm}{4.\arabic{algorithm}}
\setcounter{algorithm}{0}
\begin{algorithm}
\caption{LocateBeads (\textit{pixels})}
$\mathit{beads} \leftarrow 0$\\
$\{\hat{\bolds \beta}_{\mathrm{OLS}}^{(\mathit{beads})},\mathit{IC}^{(\mathit{beads})}\}
\leftarrow\operatorname{OLS}({\mathit{pixels},\mathit{beads}})
$\\
\textbf{repeat}\\
$
\hspace*{12pt}\cases{\mathit{beads} \leftarrow \mathit{beads}+1
\cr
\{\hat{\bolds \beta}_{\mathrm{OLS}}^{(\mathit{beads})}, \mathit{IC}^{(\mathit{beads})}\}
\leftarrow\operatorname{OLS}({\mathit{pixels},\mathit{beads}})
}
$\\
\textbf{until} $\mathit{IC}^{(\mathit{beads})}>\mathit{IC}^{(\mathit{beads}-1)}$\\
$
\mathit{beads} \leftarrow \mathit{beads}-1
$\\
$
\{\hat{\bolds \beta}_{\mathrm{MLE}}^{(\mathit{beads})}, \ell_n(\hat{\bolds \beta
}_{\mathrm{MLE}}^{(\mathit{beads})})\} \leftarrow\operatorname{MLE}({\mathit{pixels},\mathit{beads}, \hat{\bolds
\beta}_{\mathrm{OLS}}^{(\mathit{beads})})}
$\\
\textbf{repeat}\\
$
\hspace*{12pt}\cases{\mathit{beads} \leftarrow \mathit{beads}-1
\cr
\{\hat{\bolds \beta}_{\mathrm{MLE}}^{(\mathit{beads})}, \ell_n(\hat{\bolds \beta
}_{\mathrm{MLE}}^{(\mathit{beads})})\} \leftarrow\operatorname{MLE}({\mathit{pixels},\mathit{beads}, \hat{\bolds
\beta}_{\mathrm{OLS}}^{(\mathit{beads})}})
\cr
G^2_{(\mathit{beads})} \leftarrow-2\{\ell_n(\hat{\bolds \beta
}_{\mathrm{MLE}}^{(\mathit{beads})})-\ell_n(\hat{\bolds \beta}_{\mathrm{MLE}}^{(\mathit{beads}+1)})\}
}
$\\
\textbf{until} $G^2_{(\mathit{beads})}>7.81 = \chi^2_{0.95,3}$\\
\textbf{return} (${\hat{\bolds \beta}_{\mathrm{MLE}}^{(\mathit{beads}+1)}}$)
\end{algorithm}

%s5 ###
\section{Simulated examples}
\label{sec:sim}

In this section we present a series of simulated examples. We first
apply our procedure to a typical image simulated from the Poisson plus
Gaussian model presented above. Then we examine the robustness of our
procedure by applying it in three atypical scenarios: dim beads, beads
in close proximity, and beads that are not entirely contained by the
image. Finally, we investigate the sensitivity of our procedure to
misspecification of the instrumentation error.

First, we fit an image with fifteen roughly even-spaced fluorophores to
verify that our method can handle the substantial numbers that are
sometimes found in experimental data. The parameter estimates and their
approximate standard errors appear in Table~\ref{table:typ}. While the
time to fit such a larger example is considerable, the method works
well and finds the correct number of beads without difficulties.

%t2 ###
\begin{table}
\tablewidth=200pt
\caption{Localization for a Typical FIONA Image. $A=15{,}000$ for each
bead, and $S=200$, $B=200$, $\theta=100$. No estimation error for $A$
was greater than $105$, and every approximate confidence interval save
one covered the truth. The estimation errors for $S$, $B$, and $\theta$
were $0.1$, $-0.1$, and $1.7$, respectively, and their confidence
intervals covered the true values}\label{table:typ}
\begin{tabular*}{200pt}{@{\extracolsep{4in minus 4in}}ld{6.0}d{2.2}c@{}}
\hline
\textbf{Parameter}&\multicolumn{1}{@{}c@{}}{\textbf{Truth}}&\multicolumn{1}{c}{\textbf{Est error}}&$\widehat{\mathbf{SE}}$\\
\hline
$x_0$ & 23{,}566& 0.9 & 0.422\\
$y_0$ &4852 & 0.9 & 0.423\\
$x_1$ & 2522& -0.6 & 0.423\\
$y_1$ & 18{,}672& 0.1 & 0.421\\
$x_2$ &10{,}475 & -0.6 & 0.424\\
$y_2$ &4858 & -0.05 & 0.423\\
$x_3$ & 16{,}643& -0.3 & 0.422\\
$y_3$ & 19{,}505& 0.6 & 0.423\\
$x_4$ & 6842& 0.4 & 0.423\\
$y_4$ & 16{,}060& 0.1 & 0.421\\
$x_5$ &17{,}753 & -0.3 & 0.421\\
$y_5$ & 28{,}518& 1 & 0.423\\
$x_6$ & 28{,}956& 0.2 & 0.421\\
$y_6$ &6771 & 0.2 & 0.421\\
$x_7$ &27{,}512 & 0.3 & 0.419\\
$y_7$ & 3454& -0.4 & 0.419\\
$x_8$ &4165 & 0.3 & 0.422\\
$y_8$ &13{,}466 & -0.5 & 0.422\\
$x_9$ & 28{,}960& 0.4 & 0.421\\
$y_9$ &11{,}712 & -0.5 & 0.421\\
$x_{10}$ & 25{,}394& 0.1 & 0.422\\
$y_{10}$ & 28{,}468& -0.4 & 0.421\\
$x_{11}$ &29{,}112 & 0.1 & 0.423\\
$y_{11}$ &28{,}770 & -0.6 & 0.422\\
$x_{12}$ &18{,}028 & 0 & 0.422\\
$y_{12}$ &21{,}796 & 0.1 & 0.422\\
$x_{13}$ &28{,}318 & 0.2 & 0.434\\
$y_{13}$ &12{,}251 & 0.2 & 0.428\\
$x_{14}$ &27{,}757 & 0.8 & 0.432\\
$y_{14}$ &11{,}937 & 0.5 & 0.426\\
\hline
\end{tabular*}
\end{table}

\begin{figure}

\includegraphics{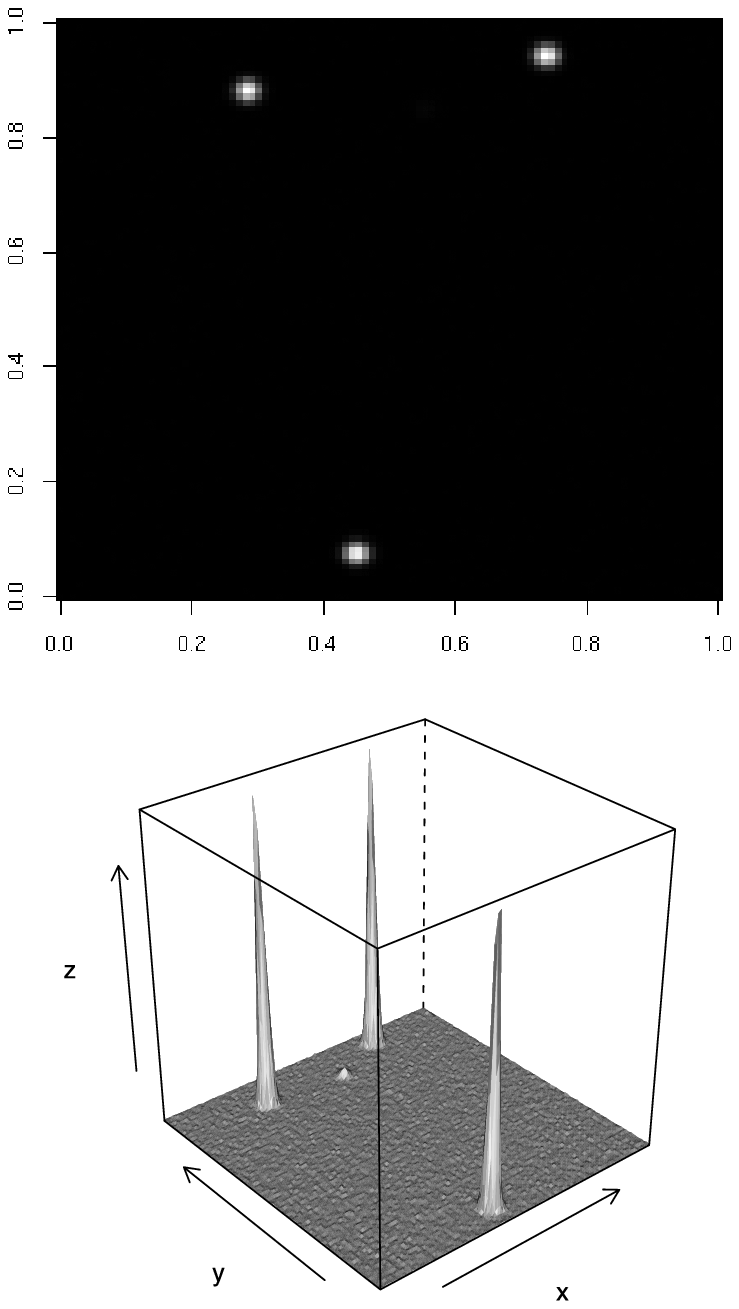}

\caption{An image with four beads, one very dim. Our procedure locates
all four beads.}
\label{fig:dim}
\end{figure}

Figure~\ref{fig:dim} shows an image that contains a bead that is very
dim ($A=400$) relative to the image's other three beads ($A=15{,}000$). We
simulated 1000 such images and applied our procedure to each. Our
procedure was able to estimate the bead's location to within a standard
error of less than six nm, which, relative to the other beads,
represents a tenfold decrease in resolution for a fortyfold decrease in
brightness. Table~\ref{table:dim} gives the results for this simulation
study. Additionally, simulations showed our algorithm capable of
consistently locating (against a background of 200) beads as dim as
$A=75$, which implies a contrast ratio, that is, the ratio of the
brightest pixel value and the background value, equal to 1.4 (versus 75
for a typical bead).

%f2 ###

%t3 ###
\begin{table}
\caption{Localization of a dim bead}\label{table:dim}
\begin{tabular*}{220pt}{@{\extracolsep{4in minus 4in}}ld{6.0}d{2.3}d{3.3}d{2.3}@{}}
\hline
\textbf{Parameter}&\multicolumn{1}{c}{\textbf{Truth}}&\multicolumn{1}{c}{\textbf{Est error}}&\multicolumn{1}{c}{\textbf{Sim SE}}&\multicolumn{1}{c@{}}{$\widehat{\mathbf{SE}}$}\\
\hline
$x_0$ & 4021 & 0.03 & 0.431 & 0.422\\
$y_0$ & 5172 & -0.13 & 0.440 & 0.422\\
$A_0$ & 15{,}000 & 69.9 & 63.339 & 49.337\\
$x_1$ & 1497 & -0.49 & 0.430 & 0.422\\
$y_1$ & 9241 & -0.04 & 0.428 & 0.422\\
$A_1$ & 15{,}000 & -8 & 61.846 & 49.508\\
$x_2$ & 7920 & -0.61 & 0.425 & 0.423\\
$y_2$ & 1807 & 0.29 & 0.405 & 0.424\\
$A_2$ & 15{,}000 & -7.3 & 60.757 & 49.513\\
$x_3$ & 6000 & 4.69 & 5.10 & 5.03\\
$y_3$ & 8722 & 11.89 & 5.30 & 5.08\\
$A_3$ & 400 & -8.761 & 12.377 & 11.11\\[5pt]
$S$ & 200 & -0.174 & 0.220 & 0.197\\
$B$ & 200 & -0.194 & 0.176 & 0.174\\
$\theta$ & 100 & 2.575 & 4.076 & 4.25\\
\hline
\end{tabular*}
\end{table}

\begin{figure}

\includegraphics{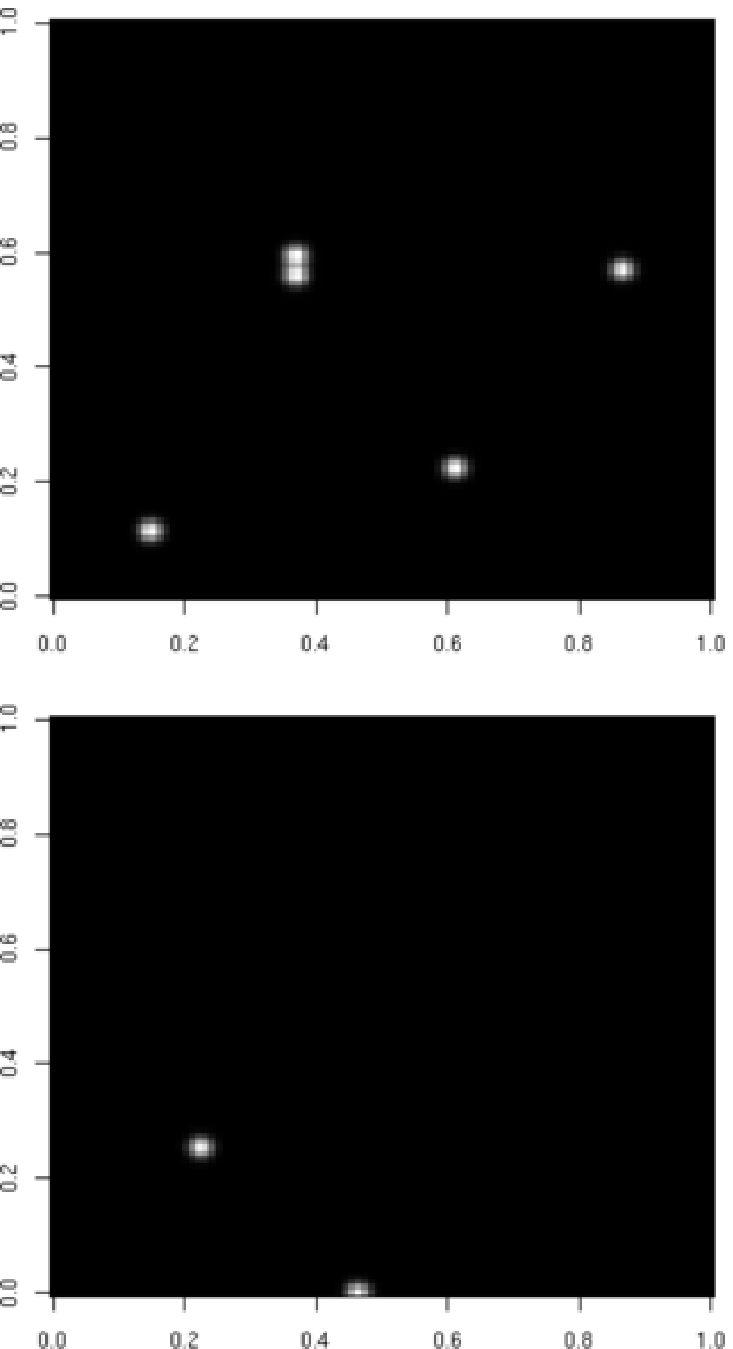}

\caption{The upper image shows two beads in close proximity. The lower
is an image with a partial bead.}
\label{fig:hangag}
\end{figure}

Figure~\ref{fig:hangag} shows an image with two beads whose centers are
separated by only 400 nm. We again applied our procedure to 1000 like
images, each image having four beads. Our algorithm was able to
distinguish the two close beads with only a slight loss of precision in
the direction of the line between the beads, as is shown in Table~\ref
{table:ag}.

The second image in Figure~\ref{fig:hangag} shows a bead whose center
is only 50 nm from the image's edge. Our algorithm was able to localize
such beads with a loss of precision in the $y$ direction that is quite
acceptable and perhaps even surprisingly small given that nearly half
of the bead is missing. Table~\ref{table:hang} reports the results for
a 1000-image study, where again each image had four beads.

%f3 ###

%t4 ###
\begin{table}
\caption{Localization of beads in close proximity}\label{table:ag}
\begin{tabular*}{220pt}{@{\extracolsep{4in minus 4in}}ld{6.0}d{2.3}d{2.3}d{2.3}@{}}
\hline
\textbf{Parameter}&\multicolumn{1}{c}{\textbf{Truth}}&\multicolumn{1}{c}{\textbf{Est error}}&\multicolumn{1}{c}{\textbf{Sim SE}}&\multicolumn{1}{c@{}}{$\widehat{{\mathbf{SE}}}$}\\
\hline
$x_0$ & 4021 & -0.22 & 0.430 & 0.422\\
$y_0$ & 5172 & 0.45 & 0.418 & 0.421\\
$A_0$ & 15{,}000 & 22.5 & 66.827 & 54.260\\
$x_1$ & 1497 & 0.06 & 0.424 & 0.422\\
$y_1$ & 9241 & 0.36 & 0.417 & 0.423\\
$A_1$ & 15{,}000 & 30 & 62.375 & 48.019\\
$x_2$ & 7920 & -0.21 & 0.467 & 0.445\\
$y_2$ & 1807 & 1.89 & 0.579 & 0.555\\
$A_2$ & 15{,}000 & -17.5 & 61.06 & 48.328\\
$x_3$ & 7920 & 0.06 & 0.449 & 0.445\\
$y_3$ & 2207 & -0.85 & 0.563 & 0.552\\
$A_3$ & 15{,}000 & 13.8 & 67.593 & 54.291\\[5pt]
$S$ & 200 & -0.119 & 0.190 & 0.178\\
$B$ & 200 & -0.196 & 0.169 & 0.173\\
$\theta$ & 100 & 6.518 & 4.261 & 4.182\\
\hline
\end{tabular*}
\end{table}

%t5 ###
\begin{table}[b]
\caption{Localization of a partial bead}\label{table:hang}
\begin{tabular*}{210pt}{@{\extracolsep{4in minus 4in}}ld{5.0}d{3.3}d{2.3}d{2.3}@{}}
\hline
\textbf{Parameter}&\multicolumn{1}{c}{\textbf{Truth}}&\multicolumn{1}{c}{\textbf{Est error}}
&\multicolumn{1}{c}{\textbf{Sim SE}}&\multicolumn{1}{c@{}}{$\widehat{\mathbf{SE}}$}\\
\hline
$x_0$ & 4021 & -0.09 & 0.413 & 0.424\\
$y_0$ & 5172 & -0.03 & 0.410 & 0.423\\
$A_0$ & 15{,}000 & 143 & 69.09 & 48.217\\
$x_1$ & 1497 & -0.39 & 0.421 & 0.424\\
$y_1$ & 9241 & 0.39 & 0.432 & 0.424\\
$A_1$ & 15{,}000 & 143 & 69.581 & 71.17\\
$x_2$ & 7920 & 0.37 & 0.430 & 0.422\\
$y_2$ & 1807 & 0.35 & 0.417 & 0.422\\
$A_2$ & 15{,}000 & 104.4 & 64.671 & 48.416\\[5pt]
$x_3$ & 6000 & 0.11 & 0.524 & 0.525\\
$y_3$ & 50 & -0.822 & 0.924 & 0.846\\
$A_3$ & 15{,}000 & 37.2 & 62.715 & 48.738\\[5pt]
$S$ & 200 & -0.51 & 0.214 & 0.187\\
$B$ & 200 & -0.119 & 0.169 & 0.175\\
$\theta$ & 100 & 0.588 & 4.243 & 4.265\\
\hline
\end{tabular*}
\end{table}

Maximum likelihood estimation is often sensitive to model
misspecification, and so we investigate the performance of our
procedure when the instrumentation error is not $N(0,\theta)$. The
instrumentation error for each image has mean zero and variance $\theta
$, but otherwise the errors are distributed rather differently. The
model was simulated, but with heavy-tailed ($t_3$ distributed) and
asymmetric (exponentially distributed) instrumentation error. More
specifically, the model was simulated according to
%
%e17 ###
\begin{equation}
Z_i\sim \operatorname{Poi}(f_i)+\sqrt{\theta/3} t_3
\end{equation}
and according to
%
%e18 ###
\begin{equation}
Z_i\sim \operatorname{Poi}(f_i)+\operatorname{Exp}\bigl(\sqrt{\theta}\bigr)-\sqrt{\theta},
\end{equation}
where $\operatorname{Poi}(\lambda)$ denotes the Poisson distribution with rate $\lambda
$, $t_\nu$ denotes the $t$ distribution with $\nu$ degrees of freedom,
and $\operatorname{Exp}(\lambda)$ denotes the exponential distribution with mean
$\lambda$. Tables~\ref{table:t} and \ref{table:exp} show that
localization was not affected by these misspecifications. Again, 1000
images were used for each study.

%t6 ###
\begin{table}
\caption{Estimation when instrumentation error is heavy-tailed}\label{table:t}
\begin{tabular*}{210pt}{@{\extracolsep{4in minus 4in}}ld{6.0}d{3.3}d{2.3}d{2.3}@{}}
\hline
\textbf{Parameter}&\multicolumn{1}{c}{\textbf{Truth}}&\multicolumn{1}{c}{\textbf{Est error}}
&\multicolumn{1}{c}{\textbf{Sim SE}}&\multicolumn{1}{c@{}}{$\widehat{\mathbf{SE}}$}\\
\hline
$x_0$ & 4021 & -0.25 & 0.433 & 0.422\\
$y_0$ & 5172 & -0.29 & 0.409 & 0.422\\
$A_0$ & 15{,}000 & -18.3 & 59.564 & 47.923\\
$x_1$ & 1497 & -0.3 & 0.437 & 0.421\\
$y_1$ & 9241 & -0.54 & 0.425 & 0.421\\
$A_1$ & 15{,}000 & 26.9 & 65.228 & 47.658\\
$x_2$ & 7920 & -0.19 & 0.428 & 0.422\\
$y_2$ & 1807 & -0.13 & 0.422 & 0.422\\
$A_2$ & 15{,}000 & 33 & 58.357 & 47.671\\
$x_3$ & 6000 & -0.6 & 0.432 & 0.421\\
$y_3$ & 8722 & -0.1 & 0.413 & 0.422\\
$A_3$ & 15{,}000 & -19.8 & 65.419 & 47.830\\
$S$ & 200 & 0.073 & 0.190 & 0.171\\
$B$ & 200 & -0.017 & 0.175 & 0.173\\
$\theta$ & 100 & 6.435 & 9.759 & 4.191\\
\hline
\end{tabular*}
\end{table}

%t7 ###
\begin{table}[b]
\caption{Estimation when instrumentation error is asymmetric}\label{table:exp}
\begin{tabular*}{210pt}{@{\extracolsep{4in minus 4in}}ld{6.0}d{4.3}d{2.3}d{2.3}@{}}
\hline
\textbf{Parameter}&\multicolumn{1}{c}{\textbf{Truth}}&\multicolumn{1}{c}{\textbf{Est error}}
&\multicolumn{1}{c}{\textbf{Sim SE}}&\multicolumn{1}{c@{}}{$\widehat{\mathbf{SE}}$}\\
\hline
$x_0$ & 4021 & -0.26 & 0.428 & 0.420\\
$y_0$ & 5172 & -0.75 & 0.398 & 0.421\\
$A_0$ & 15{,}000 & 5.1 & 60.595 & 47.619\\
$x_1$ & 1497 & -0.39 & 0.410 & 0.421\\
$y_1$ & 9241 & 0.07 & 0.422 & 0.422\\
$A_1$ & 15{,}000 & -149.7 & 62.130 & 48.146\\
$x_2$ & 7920 & -0.17 & 0.413 & 0.421\\
$y_2$ & 1807 & 0.08 & 0.442 & 0.421\\
$A_2$ & 15{,}000 & -68 & 57.774 & 47.940\\
$x_3$ & 6000 & 0.67 & 0.424 & 0.420\\
$y_3$ & 8722 & 0.04 & 0.420 & 0.419\\
$A_3$ & 15{,}000 & -74 & 61.813 & 48.109\\
$S$ & 200 & 0.12 & 0.188 & 0.170\\
$B$ & 200 & 0.44 & 0.178 & 0.173\\
$\theta$ & 100 & 5.075 & 4.798 & 4.203\\
\hline
\end{tabular*}
\end{table}

Table~\ref{table:cover} gives the coverage rates of our approximate
95\% confidence regions for all of the previously mentioned simulation
studies and for (first row) a study of 1000 typical images. The
coverage rates clearly suffer a bit for all except the typical and
asymmetric scenarios.

%t8 ###
\begin{table}
\caption{Coverage rates of approximate 95\% confidence ellipses}\label{table:cover}
\begin{tabular*}{200pt}{@{\extracolsep{4in minus 4in}}lcc@{}}
\hline
\textbf{Scenario}&\textbf{Bead type}&\textbf{Coverage rate}\\
\hline
Typical
& typical & 94.4\%\\
& typical & 95.6\%\\
& typical & 95.1\%\\
& typical & 95.4\%\\[5pt]
Dim
& typical & 95.6\%\\
& typical & 94.6\%\\
& typical & 95.2\%\\
& dim & 93.7\%\\[5pt]
Close
& typical & 95.1\%\\
& typical & 94.6\%\\
& close & 94.3\%\\
& close & 93.8\%\\[5pt]
Partial
& typical & 95.3\%\\
& typical & 94.5\%\\
& typical & 95.8\%\\
& partial & 93.0\%\\[5pt]
Heavy-tailed
& typical & 94.9\%\\
& typical & 94.3\%\\
& typical & 94.8\%\\
& typical & 93.5\%\\[5pt]
Asymmetric
& typical & 96.1\%\\
& typical & 95.6\%\\
& typical & 95.8\%\\
& typical & 95.2\%\\
\hline
\end{tabular*}
\end{table}

%s6 ###
\section{Analysis of an experimentally observed FIONA image}
\label{sec:ex}

\begin{figure}

\includegraphics{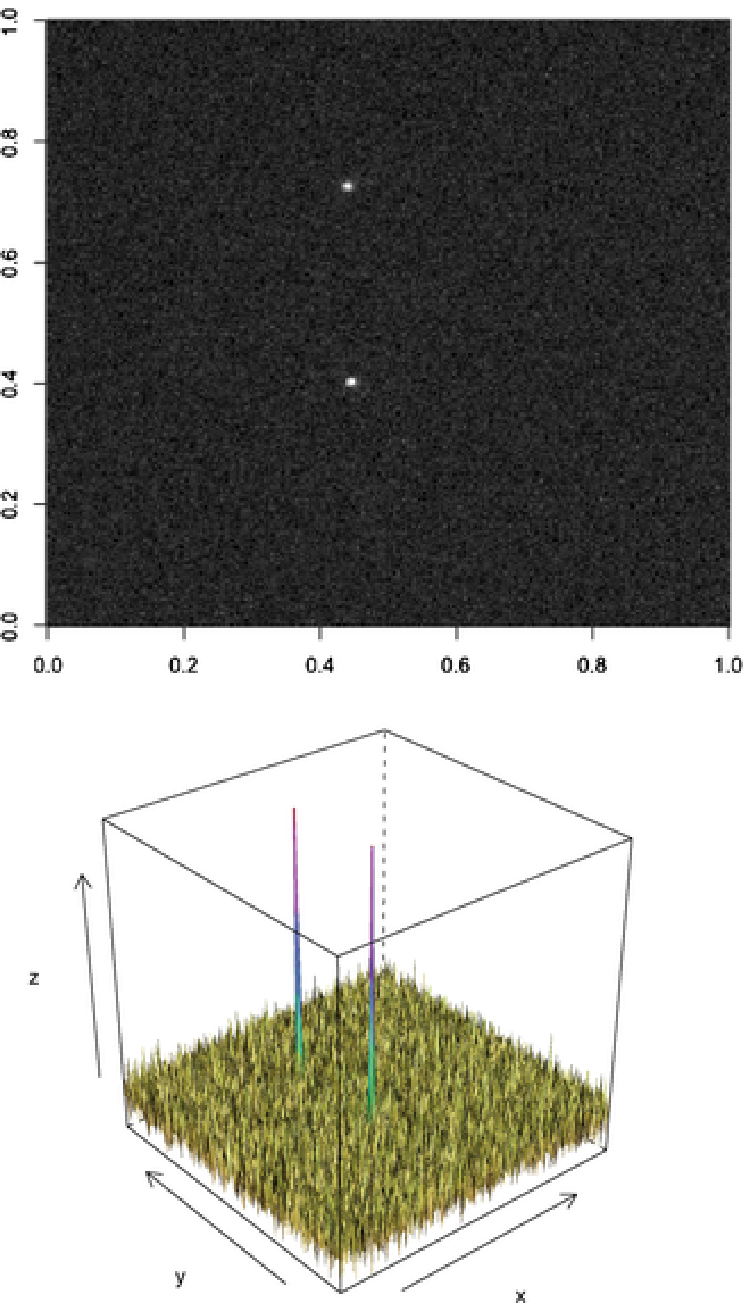}

\caption{An experimentally observed FIONA image.}
\label{fig:real}
\end{figure}

In this section we apply our procedure to an experimentally observed
FIONA image shown in Figure~\ref{fig:real}. Table~\ref{table:real}
shows our parameter estimates for this image.

%f4 ###

To evaluate the fit to the image, we ran numerous diagnostics to verify
that the observed data originates from our random field model.
Our approximate model implies that for the $i$th pixel
%
%e19 ###
\begin{equation}
Z_i\stackrel{\cdot}{\sim}N(f_i,f_i+\theta),
\end{equation}
spatially independent of the other {pixels}. If our model is correct,
then we should have, for the $i$th error,
%
%e20 ###
\begin{equation}
\varepsilon_i=\frac{Z_i-f_i}{\sqrt{f_i+\theta}}\stackrel{\cdot}{\sim}N(0,1),
\end{equation}
spatially independent of the other errors. This implies that the variogram
%
%e21 ###
\begin{equation}
\gamma(\mathbf{h})=\tfrac{1}{2}\operatorname{Var}\{\varepsilon(\mathbf{s}+\mathbf{h})-\varepsilon
(\mathbf{s})\}
\end{equation}
should equal one for all locations $\mathbf{s}$ and lag (displacement)
vectors $\mathbf{h}$.

%t9 ###
\begin{table}
\tablewidth=170pt
\caption{Parameter estimates for an experimentally observed FIONA image}\label{table:real}
\begin{tabular*}{170pt}{@{\extracolsep{4in minus 4in}}ld{6.1}d{2.4}@{}}
\hline
\textbf{Parameter}&\multicolumn{1}{c}{\textbf{Estimate}}&\multicolumn{1}{c@{}}{$\widehat{\mathbf{SE}}$}\\
\hline
$x_0$ & 12{,}168.1 & 4.83\\
$y_0$ & 7570.4 & 4.79\\
$A_0$ & 269.8 & 10.5\\
$x_1$ & 12{,}296.1 & 4.71\\
$y_1$ & 16{,}509 & 4.63\\
$A_1$ & 275.7 & 10.3\\
$S$ & 175.7 & 3.31\\
$B$ & 33.4 & 0.0418\\
$\theta$ & 80.8 & 0.632\\
\hline
\end{tabular*}\vspace*{-15pt}
\end{table}

\begin{figure}

\includegraphics{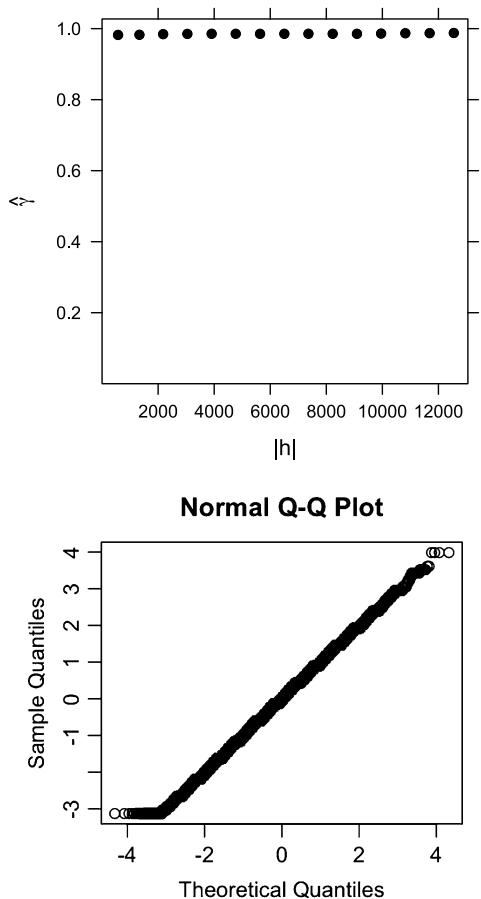}

\caption{A variogram plot and a normal probability plot of the
standardized residuals for the real FIONA image. The lags for the
variogram plot are given in nanometers, and the image in question was
approximately 27,000 nm on a side.}
\label{fig:diag}\vspace*{-10pt}
\end{figure}

We plot empirical (residual) variograms to determine spatial
independence and use a normal probability plot to check for normality
[\citet{Cressie1980Robust-estimati}]. The plots for our example image
are shown in Figure~\ref{fig:diag}. Except for some anomalous features
in the lower tail of the probability plots, the diagnostics give a good
indication that our proposed model is sound. The standardized residuals
were also checked and no blatant violations of what would be expected
for independent, identically distributed data were found.

%f5 ###

%s7 ###
\section{Conclusions}
The method outlined in this paper allows for the automated analysis of
FIONA images, including the ability to select the number of
fluorophores in an image. By using a likelihood framework, the method
also allows for standard errors to be calculated simultaneously with
the estimates. The method was then verified through simulation and the
analysis of collected data. We hope that this case study will serve as
an example of applying traditional statistical theory to enhance the
analysis of nanoscale experimental methods where algorithmic approaches
have been favored.

Since this method is largely automated through model selection
techniques, it can handle the analysis of ``movies'' by processing each
frame. Since the method also returns standard errors for the locations
of the fluorophores, this opens the possibility of creating tracking
methods to follow dynamic specimens using not only the position data
but the information on observational errors which are given.

Estimation and model selection can be done using our free C++
application, \textsf{beads}, which makes extensive use of the GNU
Scientific Library, and fit diagnostics can be carried out using our
free R software package, \textsf{FIONAdiag} [\citet{Mark-Galassi2008GNU-Scientific-}].

\printaddresses

\end{document}